\documentclass[11pt,twoside]{article}   
\usepackage{epsfig}
\usepackage{amsmath}
\usepackage{amssymb}
\date{}
\begin{document}           

\begin{center}{\bf Yilmaz Theory of SNe 1a Redshift}
\bigskip

Stanley L. Robertson \footnote{Physics Dept., Emeritus, Southwestern Oklahoma State University\\
Weatherford, OK 73096, (stan.robertson@swosu.edu)}

\begin{abstract} A redshift-luminosity distance relation in
excellent agreement with observations is calculated here for SNe
1a using the Yilmaz gravitational theory. In contrast to the
current conventional explanation based on general relativity, the
Yilmaz theory does not require a cosmological constant term that
implies the existence of ``dark energy". The Yilmaz theory
requires only one parameter; a mean mass-energy density of the
cosmos. The required value is essentially the same as the critical
density for a Friedmann-Robertson-Walker cosmological metric. The
Yilmaz theory therefore still requires the existence of
non-baryonic dark matter.
\end{abstract}
\end{center}


\section{Introduction}
The redshifts exhibited by distant SNe 1a can be encompassed by
the addition of a cosmological constant to the field equations of
general relativity. These can be written as
\begin{equation}
G_i^j = -(8\pi G/c^4)T_i^j - \delta_i^j\Lambda/c^2
\end{equation} 
where $G_i^j$ is the Einstein tensor (not to be confused with
Newton's gravitational constant in the right member) and $T_i^j$
is the matter energy-momentum tensor. $\Lambda$ is Einstein's
cosmological constant. $\Lambda c^2/(8\pi G)$ can be interpreted
as a constant energy density of the cosmological vacuum. It is a
constant ``dark energy" density that uniformly pervades
cosmological spacetime. With the adoption of a
Friedmann-Robertson-Walker (FRW) metric and an appropriate value
for $\Lambda$, the solutions of Eq. 1 provide a good description
of an expanding and accelerating universe. Fitting the solution to
the luminosity distance vs. redshift data for SNe 1a yields a
numerical value for $\Lambda$ that implies that ``dark energy"
must consitute about 73\% of the mass-energy density of the
universe (Suzuki et al. 2012). About 23\% of the remaining
mass-energy must consist of ``non-baryonic dark matter", which
leaves about 4\% ordinary baryonic matter.

If ``dark energy" represents the ground state oscillations of all
of the quantum fields within the cosmos, we might expect its value
to be roughly 120 orders of magnitude larger (Carroll 2004, Sec.
4.5) than the $\sim 10^{-8}$ erg cm$^{-3}$ that is needed to
explain the cosmological redshift observations. In addition to
this rather glaring discrepancy, the energy density of matter
would decrease in an expanding universe, which would allow only
one coincidental moment of time in which matter and vacuum energy
densities might be of comparable magnitude. Thereafter the
expansion of the universe must accelerate. In view of the
discomfort caused by the size discrepancy and the coincidence
problem, there is reason to consider a different approach to
understanding the cosmological redshifts.
\\
\section{Yilmaz Theory}
Though well established in the research literature, (Yilmaz 1971,
1982, 1992, Mizobuchi 1985) the Yilmaz gravity theory is neither
well known nor widely utilized, but it has passed all of the
previously known observational tests and remains as a viable
gravity theory. The Yilmaz theory is a metric theory of gravity in
a preferred set of coordinates. It differs from Einstein's general
relativity primarily by the role of the metric coefficients,
$g_{ij}$.

In the Einstein theory, the metric coefficients are
generalizations of the gravitational potential of Newton's theory.
In the Yilmaz theory, $g_{ij}$ is merely the generalization of
$\eta_{i j}$, the Lorentzian metric of special relativity. The
gravitational potential, $\phi$ is also extended to become a
tensor field, $\phi_\alpha ^\beta (x^k)$ such that the curved
spacetime metric is a function of the $\phi_\alpha ^\beta$; i.e.
\begin{equation}
g_{ij}=g_{ij}(\phi_\alpha ^\beta (x^k)).
\end{equation} 
In this way, gravity is regarded as a field in its own right, and
not something manifest solely by spacetime curvature. In general,
$\phi_\alpha^\beta \neq 0$ in matter free space and also
$(\partial_\nu \phi_\alpha^\beta)^2 \neq 0$ in general. Thus there
will be localized field energy in matter free space. It is this
field energy that takes the place of the cosmological constant and
permits an accurate accounting of the redshifts of SNe 1a. It is
of second degree in the derivatives of gravitational potentials
and automatically of the appropriate sign and magnitude (see
Appendix). It also contributes to spacetime curvature in a way
that was forbidden by fiat in Einstein's theory.

In the Yilmaz theory, the presence of free fall is indicated by a
locally Minkowskian metric. Removal of constraining forces that
prevent free fall is represented by choice of reference position
for potentials. Thus, unlike general relativity, free-fall is not
achieved by a coordinate transformation, but rather by the choice
of reference position for potential, such that $\phi \rightarrow
\phi+C \rightarrow 0$ and $g_{ij} \rightarrow 1$ resulting in a
locally Minkowskian metric.

Particles of mass-energy are the sources of the potentials. The
metric coefficients are functions of the potentials; i.e,
$g_{ij}=g_{ij}(\phi_\mu^\nu (x^k))$ and the potentials obey
generalized coordinates d'Alembertian equations. For a parcel of
proper mass density $\rho$ and speed $u_\mu$, $T_\mu^\nu=\rho
u_\mu u^\nu$ and:
\begin{equation}
\Box^2 \phi_\mu^\nu = (4\pi G /c^4) T_\mu^\nu
\end{equation} 
where $\Box^2=(\sqrt{-g})^{-1} \partial_i (\sqrt{-g}
~g^{ij}\partial_j)$ is the d'Alembertian operator. It is apparent
that where spacetime curvature is negligible, these equations
reduce to those of a special relativistic field theory.

One of the central tenets of the Yilmaz theory is that the speed
of light in space free of matter must be isotropic. This condition
can be enforced in part by adhering to the use of ``harmonic
coordinates" for which
\begin{equation}
\partial_i(\sqrt{-g}~g^{ij}) =0
\end{equation} 
It ensures that the phase speed of light will be the same in every
direction\footnote{A plane wave of the form $\psi = e^{i(\omega t
-{\bf k . r})}$)propagating in the space of Eq. 7 should satisfy a
generalized d'Alembertian equation, $\Box^2 \psi =
(1/\sqrt{-g})\partial_i{(\sqrt{-g}~g^{ij}\partial_j \psi)} =0$.
This will generate nonzero terms of the form {$\psi k_j
\partial_i (\sqrt{-g}~g^{ij})$} and make the phase speed of light depend
on its direction of travel unless $\partial_i(\sqrt{(-g}~g^{ij})
=0$. This harmonic gauge condition is assumed to hold for the
metric of Eq. 7 and must also hold if $\psi$ were to represent a
gravitational wave.}.

In his first presentation of the theory in terms of tensor
potentials, in 1971, Yilmaz stated the relation between the fields
$\phi_\mu^\nu$ and the metric as a functional differential
equation
\begin{equation}
dg_{\mu \nu}=2(g_{\mu \nu}d\phi - g_{\mu \alpha}d\phi_\nu^\alpha
-g_{\nu \alpha}d\phi_\mu^\alpha)
\end{equation} 
where $\phi$ is the trace of $\phi_\mu^\nu$. It was later shown
(Yilmaz 1975, 1992) that this can be integrated to yield a metric
form which is exact for many cases of physical interest:
\begin{equation}
g_{\mu \nu}=(\hat{\eta} e^{[2(\phi \hat{I}-2\hat{\phi})]})_{\mu
\nu}
\end{equation} 
where $\hat{\eta}$ is the metric of the Minkowskian background,
$\phi=\phi_k^k$ is the trace of $\hat{\phi}=\phi_\mu^\nu$ and
$\hat{I}$ is the identity matrix. The exponential function is
defined in terms of its ordered Taylor expansion and $\hat{\eta}$
is the Minkowskian metric. (In rectangular coordinates,
$\hat{\eta}$ is diagonal with elements (1, -1, -1, -1).)

Although the equations of the Yilmaz theory are decidedly
nonlinear, they are very easy to use in weak field situations. One
of their striking features is that gravitating masses move in
potentials that obey a superposition principle and they interact
via a field stress-energy tensor, $t_i^j$. Thus multiple body
interactions are easily encompassed in the Yilmaz theory (Yilmaz
1992, 1994). In contrast there are no easy multiple body solutions
in general relativity.

For a metric in coordinates $(t,x,y,z)$ one would normally need at
least the potentials $\phi_0^0, \phi_1^1, \phi_2^2$, and
$\phi_3^3$, however, for a cosmological description, we will
assume that the metric has the isotropic form:
\begin{equation}
ds^2 = e^\nu c^2 dt^2 - e^\lambda (dx^2 + dy^2 + dz^2)
\end{equation}
for which $g_{00}=e^{\nu}$ and $g_{ii} = -e^{\lambda}$.

In this case, we find from Eq. 6 that $\phi_2^2=\phi_3^3=\phi_1^1$
is required and
\begin{equation}
\nu(t,x,y,z)=6\phi_1^1-2\phi_0^0
\end{equation} 
and
\begin{equation}
\lambda(t,x,y,z)=2\phi_1^1 + 2\phi_0^0
\end{equation} 

The harmonic coordinate condition, Eq. 4, provides two relations
\begin{equation}
\partial_0 e^{(3\lambda - \nu)/2)} = \partial_0 e^{4\phi_0^0}=0
\end{equation} 
\begin{equation}
\partial_i e^{(\lambda + \nu)/2)} = \partial_i e^{4\phi_1^1}=0
\end{equation} 

It is apparent from Eqs. 10 \& 11 that we must have $\phi_0^0 =
\phi_0^0(x,y,z)$, independent of time, and $\phi_1^1=\phi_1^1(t)$,
must be independent of $x,y,z$. Eq. 3 then gives the equations to
be solved for these potentials as:
\begin{equation}
\nabla^2 \phi_0^0= - (4\pi G /c^4)\rho u_0u^0
e^{(2\phi_0^0+2\phi_1^1)}
\end{equation} 
and
\begin{equation}
 e^{-\nu} \ddot \phi_1^1 =(4\pi G/c^4)T_1^1
\end{equation} 

As Eq. 12 stands, it is inconsistent because $\phi_0^0$ is
supposed to have no time dependence, yet $\phi_1^1$ has time
dependence in the right member. This inconsistency can be removed
by choosing our observation point to be located at $r=0$, $t=0$,
where the present value of mass-energy density would be $\rho_0$.
and requiring that
\begin{equation}
\rho=\rho_0 e^{-2\phi_1^1}
\end{equation} 
Although $\rho_0=\rho_0(r)$ would remove the inconsistency, it
would describe an inhomogeneous universe. To avoid this, we
require $\rho_0 = constant$. This removes the inconsistency and
leaves Eq. 12 as
\begin{equation}
\nabla^2 \phi_0^0 = - (4\pi G /c^4)\rho_0u_0u^0 e^{2\phi_0^0}
\end{equation} 

\section{Cosmological Red Shifts}
In this section we will obtain a solution of Eqs. 13 \& 15 and a
relation between redshift and luminosity distance for SNe 1a for a
model universe consisting of an expanding, isotropic, spherically
symmetric, pressureless cosmic dust comprised of galaxy sized dust
particles. We will use the metric form of Eq. 7 as applied by an
observer located at $r=0$ at the present time, $t=0$. There is no
``universal" time for the universe in this approach nor is there a
scale factor for the entire universe. There is only the local time
of an observer at the origin of coordinates. For these present
conditions, $\lambda = \nu = 0$ at the observer's location and the
observer's local spacetime is Minkowskian. Photons emitted at
earlier (negative) times and at large distances, $r$, will be
detected as redshifted by $1+z = e^{-\nu/2}$. Photons emitted at
earlier times into a particular solid angle will be spread over a
larger aperture as the universe expands while they are in transit.
As a result, they will appear to have come from a more distant
source. In conventional FRW cosmology, the measured photon flux is
diminished by two factors of $(1+z)$. The individual photons
redshift by a factor (1 + z), and the photons hit the detector
less frequently, due to time dilation. In the present approach,
one of these factors is taken into account in $\nu \neq 0$. The
apparent luminosity distance will be enlarged and given by $d_L =
(1+z)r$.

For this model universe, it is assumed that motions of the dust
particles are always very small compared to the speed of light.
Thus only $T_0^0$ is nonzero. $ T_0^0 = \rho u_0 u^0 = \rho c^2$,
where $\rho$ is the average mass density of ``dust particles" in
the universe. Taking $T_1^1=0$, Eq. 13 shows that $\phi_1^1$ will
vary linearly with time. Taking $\phi_1^1 = C_1 t$, with $C_1$ a
positive integration constant, satisfies Eq. 13. Eq. 14 then
provides for a cosmos with a matter density that decreases with
time; i.e., an expanding universe.

Defining
\begin{equation}
R_0 = \sqrt{c^2/(4\pi G \rho_0)}, ~~x=r/R_0,~~ T=ct/R_0
\end{equation}
and converting $\nabla$ to spherical coordinates, Eq. 15 becomes
\begin{equation}
d^2\phi_0^0/ dx^2 +(2/x)d\phi_0^0/dx = -e^{2\phi_0^0}
\end{equation}
A low order solution can be obtained by expanding the exponential
function of the right member. Assuming that $\phi_0^0 = \Sigma a_n
x^{n+2}$, we find,
\begin{equation}
\phi_0^0 = -x^2/6 +x^4/60-x^6/687.3 + x^8/3565~ .~ .~ .
\end{equation} 
This fits well to $x \sim 1$, but numerical solutions are needed
for larger values of $x$. The numerical solutions of Eqs. 13 and
17 allow us to determine the metric functions
\begin{equation}
\lambda = 2C_1 T+2\phi_0^0
\end{equation} 
\begin{equation}
\nu=6 C_1 T - 2\phi_0^0
\end{equation} 

$C_1$ is proportional to the Hubble constant as will be seen by
considering the gravitational red shift that would be expected for
a photon emitted at some previous time and detected now at our
location $x=0,~T=0$. A null geodesic photon path taken from r to
zero and time T in the past to the present will have $ds^2 = 0$.
Thus, $e^{\nu/2}cdt = -e^{\lambda/2}dr$, where the negative sign
is taken because $r$ decreases as $t$ increases from the time of
emission to our detecting it at the present time. Substituting the
solutions for $\lambda$ and $\nu$ into this last relation and
rearranging, we obtain
\begin{equation}
\int_T^0 e^{2 C_1 T} dT = (1-e^{2C_1T})/2C_1 = -\int_x^0
e^{2\phi_0^0} dx
\end{equation} 

For large values of $x$, the integral on the right must be
evaluated numerically, but it is instructive to first consider the
expansions to lowest orders, for which we obtain
\begin{equation}
T + C_1 T^2 = -x +x^3/9
\end{equation}
To lowest order, we have $T = -x$, or $t= -r/c$, as expected. The
redshift of a photon, to lowest order, would be
\begin{equation}
z = e^{-\nu /2} - 1 \approx 3C_1 x = 3 C_1 \sqrt{4 \pi G
\rho_0}~r/c
\end{equation} 
It is now apparent that the Hubble constant, $H_0$, is given by
\begin{equation}
H_0 = 3C_1 \sqrt{4\pi G \rho_0}
\end{equation} 

By numerically solving Eq. 17 for $\phi_0^0$ and integrating
numerically, the integral on the right side of Eq. 21 is found to
have the limiting value of -2.1405 for very large $x$. For the
corresponding time, $T \rightarrow -\infty$, we find from Eq. 21
that $1/(2C_1) = 2.1405$, or $C_1=0.2336.$ This is a
self-consistent choice for $C_1$ that leaves only one free
parameter, $\rho_0$, to be chosen to fit the redshift-luminosity
data. With the value of $C_1$ now determined, the appropriate
time, $T$, for any $x$, can be found from
\begin{equation}
T=\ln{[1-2C_1 \int_0^x e^{2\phi_0^0}dx]}/2C_1
\end{equation} 

Once $T$ is known, the values of $~\phi_1^1,~\nu,~\lambda,~z$ and
$d_L=r(1+z)$ can be computed. Numerical solution data for
$C_1=0.2336$ and $\rho_0 = 1.06\times 10^{-29} g~cm^{-3}$ are
given in Table 1. This choice for $\rho_0$ was based on a Hubble
constant obtained by a least squares fit to 166 data points for $z
\leq 0.1$ that yielded $H_0 = 64.5 \pm 0.7~ km~ s^{-1}~Mpc^{-1}$.
(Data from The Supernova Cosmology Project, (Amanullah et al 2010,
Suzuki et al. 2012)) Eq. 24 was then used to calculate $\rho_0 =
1.06\times 10^{-29} g~cm^{-3}$ This provides a very good fit to
the supernova redshift data over the whole range of observed
redshifts as shown in Figure 1. The value of $\rho_0$ is
essentially the same as the critical density that would be
obtained for a FRW metric.
\begin{table}
\begin{center} \caption{Parameters of redshift and distance
calculations}

\begin{tabular}{cccccccc} \hline
x & $^a\phi_0^0(x)$  &  $\int_0^x e^{2\phi_0^0} dx$    & $T$ & z & $^bd_L (Mpc)$\\
\hline\\

0.000&0.000&0.000&0.000&0.000&0.000\\
0.100&-0.002&0.100&-0.102&0.073&350\\
0.200&-0.007&0.199&-0.209&0.150&750\\
0.300&-0.015&0.297&-0.320&0.233&1206\\
0.400&-0.026&0.393&-0.434&0.321&1722\\
0.500&-0.041&0.487&-0.552&0.414&2304\\
0.600&-0.058&0.577&-0.673&0.512&2958\\
0.700&-0.078&0.665&-0.796&0.616&3687\\
0.800&-0.100&0.748&-0.921&0.725&4498\\
0.900&-0.125&0.828&-1.047&0.838&5394\\
1.000&-0.152&0.904&-1.175&0.957&6380\\
1.100&-0.180&0.976&-1.303&1.080&7459\\
1.200&-0.211&1.043&-1.431&1.208&8637\\
1.300&-0.242&1.107&-1.559&1.339&9915\\
1.400&-0.275&1.167&-1.686&1.475&11296\\
1.500&-0.309&1.222&-1.812&1.614&12783\\
1.600&-0.344&1.275&-1.937&1.756&14377\\
1.700&-0.379&1.323&-2.061&1.901&16079\\
1.800&-0.415&1.368&-2.182&2.049&17890\\
1.900&-0.451&1.410&-2.302&2.198&19810\\
2.000&-0.487&1.450&-2.420&2.350&21840\\
3.000&-0.845&1.718&-3.473&3.896&47880\\
4.000&-1.164&1.853&-4.299&5.353&82840\\
5.000&-1.434&1.928&-4.945&6.625&124280\\
6.000&-1.662&1.973&-5.459&7.704&170250\\
7.000&-1.856&2.003&-5.878&8.618&219480\\
8.000&-2.023&2.024&-6.228&9.399&271200\\
9.000&-2.168&2.039&-6.525&10.075&324900\\
10.000&-2.296&2.050&-6.783&10.670&380400\\

 \hline
\end{tabular}\\
$^a$ All values calculated for $C_1=0.2336$. $^b$ These values
calculated for $\rho_0=1.06\times 10^{-29}~g~cm^{-3}$ are shown by
the solid line on Fig. 1.\\
\end{center}
\end{table}

\begin{figure}
\epsfig{figure=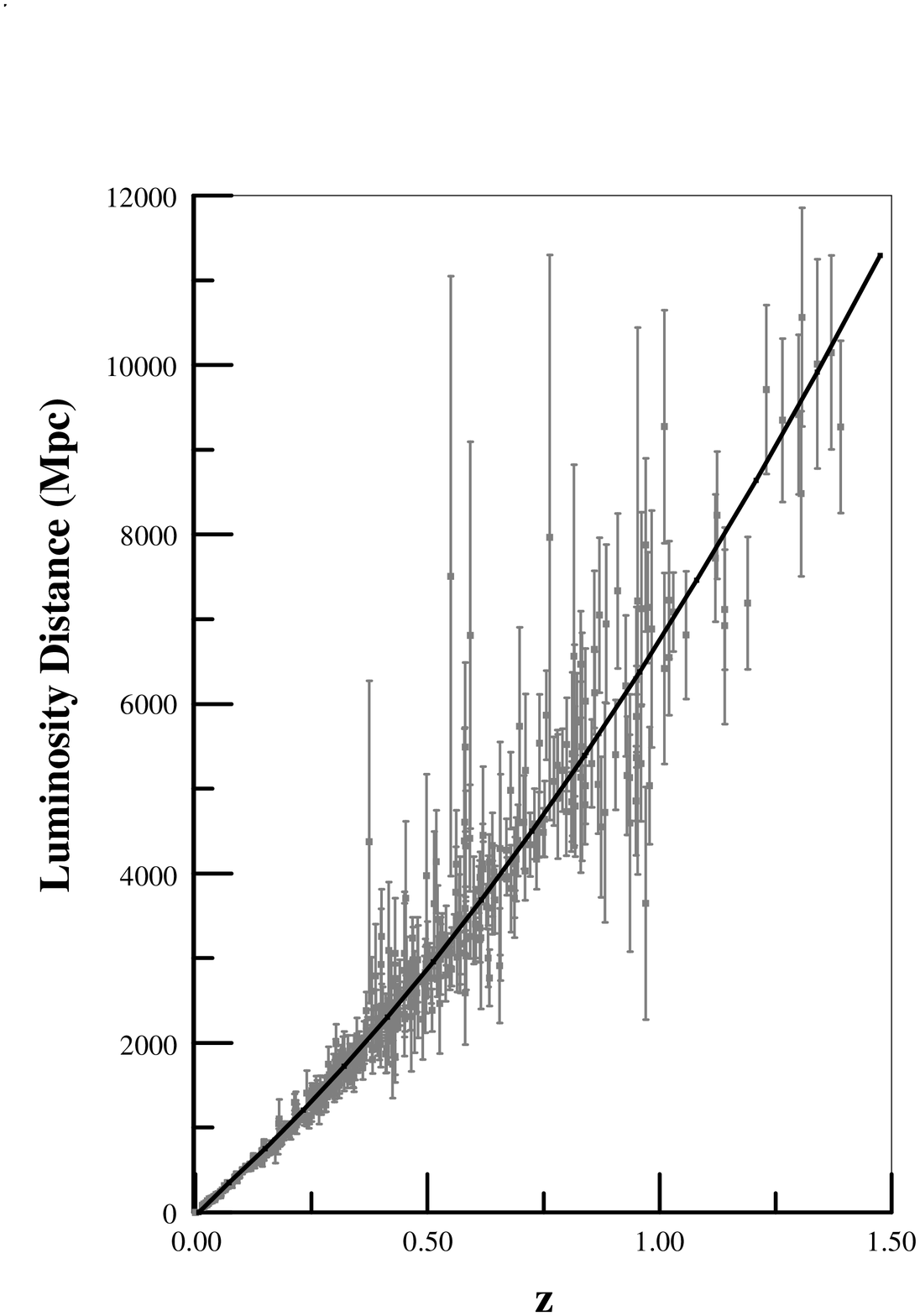, angle=0, width=16cm, height=20cm}
\caption{Luminosity distance vs redshift for SNe 1a. Data from The
Supernova Cosmology Project, (Amanullah et al 2010, Suzuki et al.
2012). The fitted curve is determined by only one free parameter,
$\rho_0 = 1.06\times 10^{-29}~ g~ cm^{-3}$ that was obtained from
the Hubble Constant fitted for $z \leq 0.1~$(see text).}
\end{figure}

\section{Discussion}
Three other attempts have been made to apply the Yilmaz theory to
cosmology. Yilmaz (1958) developed a metric for a static universe.
Increasing evidence of the inadequacy of this approach led him to
the extensions in his 1971 theory. Mizobuchi (1985) applied the
1971 theory to a cosmological model consisting of a perfect fluid.
This was not a central point of an otherwise very informative
article, but it appears to have been based on the erroneous
inclusion of a factor of $\sqrt{-g}$ in $T_\mu^\nu$, where ($-g$)
is the determinant of the metric. $T_\mu^\nu$ should have been
taken to be just the diagonal matter tensor of a perfect fluid,
$T_\mu^\nu \rightarrow (\rho c^2, -P, -P, -P)$. The approach taken
here was motivated by that of Mizobuchi (1985) but correcting the
error leads to significantly different results.

A third attempt to apply the Yilmaz theory to cosmology was
provided by Ibison (2006). Ibison assumed the correctness of the
flat-space FRW metric
\begin{equation}
ds^2= dt^2 - a(t)^2(dr^2 +r^2d\theta^2+r^2sin^2 \theta d\Phi^2)
\end{equation} 
and found a coordinate transformation to the form of Eq. 7. This
transformation was shown to satisfy the harmonic coordinate
condition, but only at the expense of leaving $\lambda$ and $\nu$
dependent only on time with no position dependence. Ibison's
transformation, $dt=a(\zeta)^3 d\zeta$, produces a result that is
equivalent to setting $\phi_0^0=0$ and is incapable of fitting the
the SNe 1a redshift data. Fig. 1 shows that the redshift data is
nicely encompassed by the Yilmaz theory and the metric of Eq. 7.

The Cosmological Principle asserts that the universe is spatially
homogenous and isotropic, but it does not demand strict adherence
to the FRW metric. The FRW metric mathematically ensures a
translational invariance that would leave the universe with the
same appearance to all observers at the same ``cosmic time", but
that is not the only way to obtain consistency with the principle.
Form invariance of Eq. 7, the requirement that $\phi_\alpha ^\beta
=0$ hold at the observer's location and the requirement that
$\rho_0 = constant$ satisfy the requirements. In this case,
however, a ``cosmic time" would have no meaning.

Yilmaz (1971) showed that his theory would change Einstein's field
equations (Eq. 1) to the form
\begin{equation}
G_i^j =-(8\pi G/c^4)(\rho u_i u^j + t_i^j)
\end{equation} 
where $t_i^j$ is a stress-energy tensor of the gravitational
field. This addition of $t_i^j$ is about the smallest correction
that one might imagine for Einstein's general relativity. Since
$t_i^j$ adds only second order corrections, leaving the first
order theory basically intact, it passes all of the weak-field
tests that have been taken as confirmation of Einstein's theory.
Without endorsing a particular form such as the Yilmaz $t_i^j$, Lo
(1995) has shown that the inclusion of a gravitational field
stress-energy term is necessary in order for Einstein's field
equations to correctly encompass his gravitational radiation
formula.

$t_i^j$ is generally of second order in the derivatives of the
potentials $\phi_\alpha^\beta$ and is small. For example, with
$x=r/R_0$ in the present calculations (see Appendix),
\begin{equation}
t_0^0= (\rho_0 c^2/2)(e^{-\nu}3C_1^2-e^{-\lambda}(\partial_x
\phi_0^0)^2)
\end{equation} 
has a value of about $8\times 10^{-10}~ erg~ cm^{-3}$ at the
observer's location at the origin of coordinates. Of course, it
has larger values at distant locations and earlier times when the
matter density of the universe was larger.

The replacement of a cosmological constant with $t_i^j$ provides a
term that is of the correct order of magnitude and sign without
regard to any other properties of the quantum vacuum. In this way,
the Yilmaz theory eliminates the need for ``dark energy", however,
the density $\rho_0=1.06\times 10^{-29} g~ cm^{-3}$, obtained here
is at least an order of magnitude larger than the known baryonic
mass density of the cosmos. Thus a need for a considerable amount
of dark matter remains in the Yilmaz theory.

Finally, it should be noted that the solution of the Yilmaz
equations found here fails to conserve energy. The conservation
law of the Yilmaz theory is the Freud identity (Yilmaz 1982,
1992). In rectangular coordinates, it would be written as
\begin{equation}
\partial_i(\sqrt{-g}~T_k^i)=0
\end{equation} 
With $T_k^i=T_0^0 = \rho_0 c^2 e^{-2\phi_1^1}$ and
$\sqrt{-g}=e^{6\phi_1^1 +2\phi_0^0}$, this would require
$\partial_0 \phi_1^1=0$ and energy-momentum would be conserved
only in a static universe.

\clearpage

\begin{center}{\bf Appendix}\end{center}

\section{A. Central body metric }
Since the Yilmaz theory is neither well-known nor widely used, it
seems appropriate to discuss some of its features here. While a
few of the basic concepts of the theory have been presented here,
the theory is well developed and informative discussions of
various aspects of the theory have also been provided by Alley
(1995), Menzel (1976), Mizobuchi (1985) and Yilmaz (1975). Perhaps
the most well-known idea of the theory is the exponential metric
for the space beyond a static, spherically symmetric mass, $M$. In
this case, $T_0^0=0$. $\phi_0^0 = \phi = GM/c^2r$ is the solution
of Eq. 3 and the metric is
\begin{equation}
ds^2=e^{-2\phi}c^2 dt^2 - e^{2\phi}(dx^2+dy^2+dz^2)
\end{equation} 
That there are no black holes in this metric may be the most
widely known aspect of the Yilmaz theory. Instead of an event
horizon at $r=2GM/c^2$, there is a photon orbit at this location.
In another respect of interest to astrophysics, the innermost
marginally stable orbit for a particle in orbit around $M$ occurs
at $r=5.24 GM/c^2$ rather than the $6GM/c^2$ of the Schwarzschild
metric.

This would accommodate most models of accretion disks for compact
objects. In addition, it should be noted that it might be possible
for objects to become so compact that $\phi \gg 1$. In this
circumstance, photons emitted at the surface would be extremely
red shifted as observed distantly and very little, if any,
luminosity would be observed for the object. A correct treatment
of realistic compact objects of astrophysical interest will
require consideration of the trapped radiation fields, especially
in cases involving an active collapse process.

Although it is unlikely that massive point particles actually
exist, $g_{00} = e^{-2\phi}$ would be zero only for $r=0$. This
should be regarded as just a classical physics point particle
singularity rather than an event horizon. The Kretschmann
invariant is zero rather than divergent at $r=0$. There is no
curvature singularity there.

\section{Gravitational field stress-energy tensor}

The Yilmaz gravitational energy expression, $t_i^j$, is
essentially Einstein's gravitational stress-energy pseudotensor
expressed in terms that eliminate pseudotensors and leave a true
tensor quantity (Yilmaz 1992, Yilmaz \& Alley 1999). First define
terms:
\begin{equation}
{\bf g}^{ij}=\sqrt{-K}g^{ij}~~~{\bf g}_{ij}=g_{ij}/\sqrt{-K}
\end{equation}
and
\begin{equation}
W_i^j=(1/8\sqrt{-K}){\bf g}^{jk}\{\bar{\partial}_k{\bf
g}_{ab}\bar{\partial}_i{\bf g}^{ab}
-2\bar{\partial}_k\sqrt{-K}\bar{\partial}_i(1/\sqrt{-K})-2\bar{\partial}_a{\bf
g}_{kb}\bar{\partial}_i{\bf g}^{ab}\}
\end{equation}
Here the overbar represents a covariant derivative with respect to
local Minkowskian coordinates which share the same origin and
orientation as those of the general metric.
$\sqrt{-K}=\sqrt{-g}/\sqrt{-\eta}$, where $\sqrt{-g}$ is the
determinant of the metric and $\sqrt{-\eta}$ is the determinant of
the metric of the Minkowskian background. In rectangular
coordinates (x,y,z,t), $\sqrt{-\eta}=1$ and all Christoffel
symbols vanish, leaving a normal partial derivative.

This elaborate derivative procedure is necessary to eliminate
pseudotensors that might otherwise arise. The pseudotensor problem
has been discussed in detail (Yilmaz 1992). This procedure
eliminates them, but it is merely a mathematical artifice. No
bimetric theory is intended here and there is no further need of a
Minkowskian metric. Particle motions under the influence of only
gravitational forces would follow geodesics of the general metric,
not the Minkowskian metric. Pseudotensor problems can be avoided
in two ways. The first simply consists of the use of rectangular
coordinates, in which they never appear. The second is to take
derivatives as covariant derivatives in local Minkowskian
coordinates. Preliminary considerations aside, $t_i^j$ is given as
a true tensor quantity as
\begin{equation}
t_i^j=W_i^j -(1/2)W_k^k
\end{equation}

Yilmaz often used expressions that incorporated a harmonic
coordinate condition. The expressions above for $W_i^j$ and
$t_i^j$ were based on Pauli's decomposition of the Einstein tensor
with no harmonic coordinate conditions included.

Evaluating Eq. 33 for the metric of Eq. 7 and complete spherical
symmetry for which $r$ is the only spatial variable yields
\begin{equation}
t_0^0 = - t_1^1 = (c^4/8\pi G)(e^{-\nu}3\dot{\lambda}^2/4 +
e^{-\lambda}(\lambda' \nu'/2+\lambda'^2/4))
\end{equation} 
and
\begin{equation}
t_2^2 = t_3^3 = (c^4/8\pi G)(-e^{-\nu}3\dot{\lambda}^2/4 +
e^{-\lambda}(\lambda' \nu'/2+\lambda'^2/4).
\end{equation}
Here dots represent partial derivatives with respect to time, and
primes represent partial derivatives with respect to the radial
coordinate, $r$. Then using the expressions of Eqs. 34 \& 35 for
the right member of Eq. 27, the field equations for $G_0^0,~
G_1^1$,and $G_2^2=G_3^3$ become
\begin{equation}
e^{-\lambda}[(1/r^2)\partial_r (r^2
\partial_r \lambda) +  \lambda'
(\lambda'+\nu')/2]= -(8 \pi G/c^4) T_0^0
\end{equation} 
\begin{equation}
\-e^{-\nu}(\ddot{\lambda}+ \dot{\lambda}
(3\dot{\lambda}-\dot{\nu})/2)+e^{-\lambda}(\lambda'+\nu')/r = (-8
\pi G/c^4)T_1^1
\end{equation} 
\begin{equation}
\begin{split} -e^{-\nu}(\ddot{\lambda}+
\dot{\lambda} (3\dot{\lambda}-\dot{\nu})/2)
+(1/2)e^{-\lambda}[\lambda''+\nu''+(\lambda'+\nu')/r
+(\lambda'+\nu')^2/2] = (-8 \pi G/c^4)T_2^2
\end{split}
\end{equation} 
The generalized d'Alembertian Eqs. 3 have time dependence that is
not evident in Eq. 36. If one wished to use harmonic coordinates
and the relations, $\lambda=2\phi_1^1+2\phi_0^0$ and
$\nu=6\phi_1^1-2\phi_0^0$, the Eqs. 3 would yield
\begin{equation}
\Box^2 \lambda = (8\pi G/c^4)(T_0^0+T_1^1)
\end{equation} 
and would include second time derivatives $e^{-\nu}
\ddot{\lambda}$ that are not in Eq. 36. As noted by Lo (1995), the
more general time dependence of the d'Alembertian equations is
necessary in order to encompass gravitational waves. The metric
form of Eq 6 and the d'Alembertian equations have been shown to
describe them well (Mizobuchi 1985). Further, in cases with no
time dependence, but where $T_1^1, T_2^2$ or $T_3^3$ would be
nonzero, one finds that the use of harmonic coordinates would lead
to $\lambda'+\nu'$=0 and no solution for Eqs. 37 \& 38. While Eq.
27 leads to Eqs. 36 \& 37 and these reduce to Eqs. 13 \& 17 under
the same assumed conditions, Eq. 27 must be regarded as having
only limited applicability.

\section{Quantum possibilities}
The question of how quantum mechanics and general relativity might
be reconciled has recently been sharpened by considering what
happens to a freely falling particle of matter approaching an
event horizon. The possibility that it might meet a radiative
``firewall" has recently become a very active research topic
(e.g., Abramowicz, Kluzniak \& Lasota 2013, Anastopoulos \&
Savvidou, 2014, Hawking 2014). This is a problem of such
importance that we should consider all aspects; however, the
necessity of event horizons seems not to have been questioned in
astrophysics. They have been accepted without proof. Although
there are many astronomical objects that are known to be compact
and massive enough to be black holes, if event horizons exist,
none have been shown to possess this quintessential feature of a
black hole.

Einstein developed general relativity with the aim of explaining
gravitational phenomena as manifestations of spacetime curvature
alone. In his field equations he included all forms of energy as
sources of gravitation and curvature but expressly rejected a
separate gravitational field as a source of energy. Instead of
having separate gravitational potentials, the metric coefficients
of general relativity take the dual roles of potentials and
descriptors of spacetime geometry. One of the problems that this
presents for quantum theory is that the covariant derivatives of
the metric tensor are identically zero. Potentials such as
$\phi_0^0$ and $\phi_1^1$ that exist separately from the metric
may provide a path to a quantum theory of gravity (Yilmaz 1995,
1997, 1980). This needs further exploration.

Although there have been indications of small things amiss with
general relativity, such as the failure to have a complete
correspondence limit with special relativity (Yilmaz 1975, Alley
1995)), they have not generally led to serious consideration of
rival theories. Even serious difficulties such as the failure to
encompass the quadrupole gravitational radiation formula (Wald
1984; Yu 1992, Lo 1995) have been ignored. To the contrary,
astrophysicists have stretched the applications of the theory to
the point of accepting the existence of black holes, singularities
and dark energy. In view of the ease with which the Yilmaz theory
removes these, they may not be necessary at all. While ``black
holes" have become a part of the mystique of astrophysics that may
persist as descriptive terminology even if event horizons are
abandoned, dark energy has much the same appeal as adding more
epicycles. It may soon be forgotten.

\section{Compact objects}
Removing event horizons from the astrophysical menagerie does,
however, leave a need for a new understanding of the nature of the
gravitationally collapsed and compact objects that are presently
thought by many to be black holes. The luminosity differences
between a very large redshift, $z$, and the $z = \infty$ of a
black hole might be small and subtle. Other differences, such as
the presence of magnetic fields, might betray the lack of an event
horizon. Robertson \& Leiter (2002) presented evidence for the
existence of intrinsic magnetic moments in stellar mass black hole
candidates. They later devised a magnetic, eternally collapsing
object (MECO) model that could account for the observations and
extended its application to active galactic nuclei, including Sgr
A* (Robertson \& Leiter 2003, 2004, 2006, 2010). Additional
observational evidence for magnetic moments in AGN has also been
found (Schild, Leiter \& Robertson 2006, 2008). The MECO model
needs minor revision to incorporate the Yilmaz exponential metric.

Since the objects presently considered to be black holes are too
massive and compact to be supported by neutron degeneracy
pressure, they most likely would collapse to a size that can be
supported by internal radiation pressure (Mitra 2006). They might
well become quark-gluon plasmas. At the same time, their surface
emissions must occur with such extreme redshifts that their
distantly observed luminosity would be quite low. In this regard,
the ECO (e.g., Mitra 2000-2006) or MECO (e.g., Robertson \& Leiter
2002-2006) models, which only need large gravitational redshifts
and/or intrinsic magnetic fields to function may possibly be
encompassed within the Yilmaz theory. This remains to be worked
out for spacetimes dominated by electromagnetic radiation fields.
\\

\end{document}